\documentclass[12pt]{article}
\usepackage{a4wide}
\usepackage{amsmath,amssymb,bbm}
\usepackage{multirow}
\def\a{\alpha}
\def\b{\beta}

\def\e{\epsilon}
\def\m{\mu}
\def\n{\nu}
\def\r{\rho}
\def\s{\sigma}
\def\de{\partial}

\begin{document}
\setcounter{page}{1}
%

%%%%%%%%%%%%%%%%%%%%%%%%%%%%%%%%

%

%%%%%%%%%%%%%%%%%%%%%%%%%%%%%%%%%%%%
%Without pictures use this macro
\def\pct#1{(see Fig. #1.)}
%%%%%%%%%%%%%%%%%%%%%%%%%%%%%%%%%%%
%With pictures use this macro
%\def\pct#1{\input epsf \centerline{ \epsfbox{#1.eps}}}

%%%%%%%%%%%%%%%%%%%%%%%% FRONT PAGE %%%%%%%%%%%%%%%%%%%%%%%%%%%%%%%%%%%%%
\begin{titlepage}
\hbox{\hskip 12cm KCL-MTH-10-01  \hfil}
% \hbox{\hskip 12cm hep-th/0410185 \hfil}
%\hbox{\hskip 12cm \hfil}
%\hbox{\hskip 12cm DRAFT $$Revision: 1.61 $$ \hfil}
%\hbox{\hskip 12cm $$Author: t58 $$ \hfil}
%\hbox{\hskip 12cm $$Date: 2003/10/31 14:37:00 $$ \hfil}
%\end{flushright}
\vskip 1.4cm
\begin{center}  {\Large  \bf  Local $E_{11}$ and the gauging of the trombone symmetry}

\vspace{1.8cm}

{\large \large Fabio Riccioni } \vspace{0.8cm}

{\sl Department of Mathematics\\
\vspace{0.3cm}
 King's College London  \\
\vspace{0.3cm}
Strand \ \  London \ \ WC2R 2LS \\
\vspace{0.3cm} UK} \\
\vspace{.8cm} {E-mail: {\tt Fabio.Riccioni@kcl.ac.uk}}
\end{center}
\vskip 1.5cm

\abstract{In any dimension, the positive level generators of the
very-extended Kac-Moody algebra $E_{11}$ with completely
antisymmetric spacetime indices are associated to the form fields of
the corresponding maximal supergravity. We consider the local
$E_{11}$ algebra, that is the algebra obtained enlarging these
generators of $E_{11}$ in such a way that the global $E_{11}$
symmetries are promoted to gauge symmetries. These are the gauge
symmetries of the corresponding massless maximal supergravity. We
show the existence of a new type of deformation of the local
$E_{11}$ algebra, which corresponds to the gauging of the symmetry
under rescaling of the fields. In particular, we show how the gauged
IIA theory of Howe, Lambert and West is obtained from an
eleven-dimensional group element that only depends on the eleventh
coordinate via a linear rescaling. We then show how this results in
ten dimensions in a deformed local $E_{11}$ algebra of a new type.}

\vfill
\end{titlepage}
%%%%%%%%%%%%%%%%%%%%%%%%%%%%%%%%%%%%%%%%%%%%%%%%%%%%
\makeatletter \@addtoreset{equation}{section} \makeatother
\renewcommand{\theequation}{\thesection.\arabic{equation}}
\addtolength{\baselineskip}{0.3\baselineskip}
%%%%%%%%%%%%%%%%%%%%%%%%%%%%%%%%%%%%%%%%%%%%%%%%%%%%

\section{Introduction}
Given a supergravity theory with a global internal symmetry group
and abelian vectors transforming in a representation of this group,
the gauging of a subgroup thereof consists in deforming this theory
turning on a gauge coupling, and collecting a subset of the vectors
in the adjoint representation of the gauge subgroup, compatibly with
gauge invariance with respect to the gauge subgroup and with
supersymmetry. In this paper we will only be interested in theories
with maximal supersymmetry. The first, and probably one of the best
known examples of a gauged theory with maximal supersymmetry is the
four dimensional ${\cal N}=8$ theory of \cite{dewitnicolai}, that is
a deformation of the massless maximal supergravity of
\cite{cremmerjuliaD=4} where an $SO(8)$ subgroup of the internal, or
Cremmer-Julia, symmetry group $E_{7(7)}$ is gauged (we refer to the
internal symmetry $E_{11 -D(11-D)}$ of the massless maximal
supergravity in $D$ dimensions as Cremmer-Julia \cite{sugraD=5}
symmetry). Gauged supersymmetric theories are sometimes called
massive theories because supersymmetry typically relates coupling
constants with mass terms.

A method of obtaining a lower dimensional gauged supergravity theory
starting from a massless higher dimensional one is due to Scherk and
Schwarz \cite{scherkschwarz}. If the higher dimensional theory
possesses an internal symmetry, one can perform a dimensional
reduction with the fields  depending on the internal coordinate via
a linear internal symmetry transformation proportional to a mass
parameter $m$. Because of the symmetry of the higher dimensional
theory, this procedure is bound to give a consistent lower
dimensional theory, in the sense that in the lower dimension there
is no dependence on the internal coordinate. This resulting theory
is a massive theory, with masses proportional to the parameter $m$.

As an example, we can consider the Scherk-Schwarz reduction of the
IIB theory to nine dimensions \cite{massiveD=9,fibrebundles}. The
IIB theory possesses an $SL(2 ,\mathbb{R})$ symmetry with generators
$R^i$, $i=1,2,3$. One thus performs a generalised dimensional
reduction to nine dimensions, in which the fields transform under
$SL(2 ,\mathbb{R})$ linearly in the internal coordinate and
proportionally to the mass parameter $m_i$ in the triplet of $SL(2,
\mathbb{R})$. This gives rise to a massive maximal supergravity in
nine dimensions, with mass $m_i$.

There are gauged supergravities that are not of the type discussed
so far in this introduction. These arise from the gauging of the
global scaling symmetry that leaves the field equations invariant,
but rescales the action. This symmetry is not a symmetry of the
Cremmer-Julia type, and it is referred to as ``trombone'' symmetry
(it is important to observe, though, that the trombone symmetry
plays a crucial role in understanding the occurrence of the
Cremmer-Julia symmetries in the lower dimensional theories
\cite{cjlp1}). The first example of such a theory is the gauged IIA
theory of Howe, Lambert and West \cite{hlw}. The massless IIA theory
\cite{IIA} has an internal symmetry $\mathbb{R}^+$ under shifts of
the dilaton, and one can consider a combination of this symmetry and
the scaling symmetry that leaves the vector invariant. This combined
symmetry can thus be gauged, resulting in a Higgs mechanism in which
the dilaton field is eaten by the vector, which becomes massive. The
fact that the scaling symmetry is not a symmetry of the lagrangian
implies that this theory does not admit a lagrangian formulation,
but only field equations. It is probably unnecessary to stress that
this theory is different from the massive IIA theory of Romans
\cite{romans}, corresponding to a deformation of the massless IIA in
which the vector is eaten by means of a Higgs mechanism in which the
2-form becomes massive.

In \cite{fibrebundles} the gauged IIA theory was shown to arise from
a generalised Scherk-Schwarz dimensional reduction from eleven
dimensional supergravity. This corresponds to performing a
dimensional reduction from eleven to ten dimensions in which the
fields depend on the internal coordinate in terms of a linear
rescaling. Given that the eleven-dimensional scaling symmetry is a
symmetry of the field equations, the ten dimensional equations do
not depend on the internal coordinate and as such the truncation to
ten dimensions is consistent from this point of view. The
lagrangian, though, has an overall scaling symmetry which is linear
in the internal coordinate, and thus the truncation to ten
dimensions is not consistent at the level of the lagrangian. This is
another way of seeing that the theory does not have a lagrangian
formulation.

Maximal supergravity theories have a very elegant and natural
classification in terms of the very-extended infinite-dimensional
Kac-Moody algebra $E_{11}$ \cite{E11}. This algebra was first
conjectured in \cite{E11} to be a symmetry of M-theory. The maximal
supergravity theory in $D$ dimensions corresponds to decomposing
$E_{11}$ in terms of $GL(D,\mathbb{R}) \otimes E_{11-D}$, and thus
the occurrence of the internal symmetry $E_{11-D}$ appears natural
from this perspective. In particular, the IIA theory naturally has
from the $E_{11}$ viewpoint an $\mathbb{R}^+$ symmetry that
corresponds to the shift of the dilaton. Decomposing the adjoint
representation of $E_{11}$ with respect to the subalgebra associated
to the IIA theory one obtains generators that are associated to the
IIA fields and their duals \cite{E11}. In this IIA decomposition of
the $E_{11}$ algebra there is a generator with nine antisymmetric
ten-dimensional spacetime indices, which is associated to a 9-form
in the IIA theory \cite{axeligorpeter}. This 9-form has a 10-form
field strength, which can be thought as the dual of the mass
parameter of Romans. Therefore the Romans massive IIA is naturally
encoded in $E_{11}$ \cite{igorpeterromans}.

More generally, decomposing the $E_{11}$ algebra in a given
dimension and considering only the level zero generators (that is
the generators of $GL(D,\mathbb{R}) \otimes E_{11-D}$, that are
associated to the graviton and the scalars) and the positive level
generators with completely antisymmetric indices, that are
associated to forms, one finds in all cases the field content of the
$D$-dimensional supergravity theory, in a democratic formulation in
which all fields appear together with their magnetic duals
\cite{axeligorpeter,fabiopeterE11origin}. One also finds generators
associated to $D-1$ forms, that are not propagating fields.
Remarkably, these generators are in one to one correspondence with a
constant scalar quantity, the so called embedding tensor, that
parametrises all possible gaugings of subgroups of the internal
symmetry $E_{11-D}$ in any dimension, and can be thought of as
belonging to a representation of $E_{11-D}$
\cite{nicolaisamtlebenD=3,dWSTgeneral,dWSTD=5,samtlebenweidnerD=7,dWSTmagnetic,dWSTD=4,embeddingtensorD=6},
which indeed is the same representation as the one to which the
$D-1$ form generators belong
\cite{fabiopeterE11origin,ericembeddingtensor}. Exactly like in the
case of Romans, one thinks of the $D-1$ form fields as being dual to
the embedding tensor, obtaining in this way a classification of all
possible maximal gauged supergravities in terms of $E_{11}$.

In the non-linear realisation, the action of positive level $E_{11}$
generators with completely antisymmetric spacetime indices
corresponds to gauge transformations for the associated form fields
that are linear in the spacetime coordinates, and one wants to
enlarge the algebra so that it includes arbitrary gauge
transformations. This was done in \cite{fabiopeterogievetsky}, and
the corresponding algebra includes the non-negative level generators
as well as momentum and an infinite set of additional generators,
that were called Ogievetsky, or Og generators, that correspond to an
expansion in the spacetime coordinates of the gauge parameters. This
extension is dimension-dependent, and it was called
$E_{11,D}^{local}$ in \cite{fabiopeterogievetsky}. From the
non-linear realisation of the $E_{11,D}^{local}$ algebra with as
local subalgebra the $D$ dimensional Lorentz algebra times the
maximal compact subalgebra of $E_{11-D}$ one computes all the field
strengths of the massless maximal supergravity in $D$ dimensions.

Given the local $E_{11}$ algebra in $D$ dimensions, one can consider
its possible massive deformations. In \cite{hierarchyE11} the
deformations that do not involve the $GL(D, \mathbb{R})$ generators
were studied, and the consistency of the deformed algebra implies
that all possible deformations are parametrised by a constant
quantity that turns out to be the embedding tensor. All the possible
deformations are thus in one to one correspondence with all the
possible gauged supergravities resulting from the gauging of a
subgroup of $E_{11-D}$. The non-linear realisation then provides an
extremely simple and powerful method to compute the field strengths
and gauge transformations of the fields.

If the gauged supergravity theory arises from a dimensional
reduction, this can be seen from the $E_{11}$ point of view in terms
of the fact that the deformed generators arise from a redefinition
involving the undeformed $E_{11}$ and Og generators in the higher
dimension. This was shown in detail in \cite{fabiopeterogievetsky}
for the case of the Scherk-Schwarz reduction of IIB to nine
dimensions. Taking the local $E_{11}$ group element associated to
the ten-dimensional IIB theory, the Scherk-Schwarz reduction
corresponds to transforming this group element by an $SL(2,
\mathbb{R})$ transformation that is linear in the internal
coordinate and in the mass parameter $m_i$, with the rest of the
group element only depending on the nine-dimensional coordinates.
From the nine-dimensional viewpoint, this results in an algebra that
is deformed by the mass parameter $m_i$ with respect to the algebra
associated to the massless nine-dimensional theory.

In this paper we show that the construction of
\cite{fabiopeterogievetsky} admits additional deformations that are
associated to the gauging of the trombone symmetry. In particular we
show that the gauged IIA theory of \cite{hlw} naturally arises as a
deformation of the local $E_{11}$ algebra of a new type. We show
this by considering an eleven-dimensional group element that only
depends on the eleventh coordinate by a linear scaling, while the
fields are taken to only depend on the ten-dimensional coordinates.
This exactly reproduces the generalised Scherk-Schwarz construction
of \cite{fibrebundles}. The fact that the symmetry that one is
gauging is not a symmetry of the lagrangian corresponds from this
point of view to the fact that the Maurer-Cartan form has an
explicit dependence on the eleventh coordinate. Still, there is a
very natural way of interpreting the results in ten dimensions, as
will be explained in the paper. The resulting ten-dimensional
algebra is the algebra corresponding to the IIA theory of
\cite{hlw}, and the new feature is that the deformation involves not
only the generator of the internal symmetry, that is the scalar
generator associated to the dilaton, but also the scaling generator
that is the trace of the $GL(10,\mathbb{R})$ generators. Recently a
complete classification of this type of maximal gauged
supergravities in any dimension was performed in \cite{henning}
using the embedding tensor formalism. The analysis of the
corresponding deformations of the local $E_{11}$ algebra will be
presented in a separate paper \cite{trombone2}.

It is important to observe that the local $E_{11}$ algebra is not
compatible with the full $E_{11}$ symmetry, including the negative
level generators. The approach taken in
\cite{fabiopeterogievetsky,hierarchyE11} was therefore to include
only the non-negative level generators, and from this approach
$E_{11}$ is not a symmetry of the eleven-dimensional group element.
This is the approach taken in this paper. An attempt to describe
gauged supergravity theories compatibly with the full $E_{11}$
symmetry, based on extending the momentum operator including
infinitely many charge generators to form an $E_{11}$ representation
\cite{l1multiplet}, was made in \cite{E11fivedimensions}. That
approach will not be discussed in this paper.

The paper is organised as follows. Section 2 contains a review of
the description of gravity as a non-linear realisation of
\cite{fabiopeterogievetsky}, as well as some comments on its
dimensional reduction. These results are useful for the main result
of the paper, that is the $E_{11}$ description of the gauged IIA
theory of \cite{hlw}, which is contained in section 3. Finally,
section 4 contains the conclusions.

\section{On gravity as a non-linear realisation}
In this section we first review the formulation of gravity as a
non-linear realisation of \cite{fabiopeterogievetsky}, and we then
show that deformations of this algebra correspond to field
redefinitions, and we finally discuss the issue of frame dependence
in the dimensional reduction. The aim of this section is to set up
the framework for the main result of the paper, which is contained
in the next section.

We want to describe gravity as a non-linear realisation of the
algebra of diffeomorphisms with the Lorentz algebra as local
subalgebra. This was originally achieved  in the four-dimensional
case in \cite{ogievetsky,borisovogievetsky}, where the algebra of
diffeomorphisms was realised as the closure of $IGL(4,\mathbb{R})$
with the conformal group $SO(2,4)$. This was generalised to $D$
dimensions in \cite{petersuperconformal}, where a vierbein rather
than a metric was introduced (the metric indeed arises using the
Lorentz group to make a particular choice of coset representative).

The more straightforward approach of \cite{pashnev} (see also
\cite{kirsch}) is to consider directly the algebra of
diffeomorphisms, which is the infinite dimensional algebra generated
by
  \begin{equation}
  P_\mu \quad , \qquad  K^\mu{}_\nu \quad ,\qquad  K^{\mu_1 \mu_2}{}_\nu \quad \ldots \quad  K^{\mu_1\ldots \mu_n}{}_\nu \ldots
  \end{equation}
with $K^{\mu_1\ldots \mu_n}{}_\nu=K^{(\mu_1\ldots \mu_n )}{}_\nu$,
satisfying the commutation relations
  \begin{equation}
  [ K^\m{}_\n  , P_\r ] = - \delta^\m_\r P_\n \label{KwithP}
  \end{equation}
  \begin{equation}
  [ K^{\mu_1\ldots \mu_n}{}_\nu , P_\rho ]= (n-1) \delta_\rho^{(\m_1}K^{\m_2\ldots
  \m_n)}{}_\n \qquad \quad n > 1 \label{OgnwithP}
  \end{equation}
and
  \begin{equation}
  [K^{\m_1\ldots \m_n}{}_\r, K^{\n_1\ldots \n_m}{}_\s]= (n+m-1 )\left[{1 \over n}
  \delta_\r^{(\n_1|}
  K^{\m_1\ldots \m_n|\n_2\ldots \n_m)}{}_\s - {1 \over m}
  \delta_\s^{(\m_1}K^{\m_2\ldots
  \m_n)\n_1\ldots \n_m}{}{}_\r \right] \ . \label{OgnwithOgm}
  \end{equation}
Here the $GL(D ,\mathbb{R})$ indices $\m$, $\n$, ... go from 1 to
$D$ and an upstairs index denotes the ${\bf D}$ and a downstairs
index the ${\bf \overline{D}}$ of $GL(D,\mathbb{R})$. Note that the
last equation for $n=m=1$ is the $GL(D,\mathbb{R})$ algebra. A
realisation of the algebra of eqs. (\ref{KwithP}), (\ref{OgnwithP})
and (\ref{OgnwithOgm}) can be obtained in terms of the position and
derivative operators $y^\m$ and $\de_\m = \de / \de y^\m$ by the
identification
  \begin{equation}
  P_\m = \de_\m \quad \qquad \quad K^{\m_1 \m_2...\m_n}{}_\n =
  \frac{1}{n} y^{\m_1} y^{\m_2} ... y^{\m_n} \de_\n \quad .
  \label{realisation}
  \end{equation}
One can assign a grade to the generators - that is $K^{\m_1\ldots
\m_{n}}{}_\n$ has grade $n-1$ and  $P_\m$ has grade $-1$ - which is
preserved by the algebra above. Note that the grade of a generator
is its dimension when the generator is realised in terms of position
and momentum operators as is eq. (\ref{realisation}). The generators
of grade $n$ higher than zero, that is all the generators apart from
the momentum generator $P_\m$ and the $GL(D, \mathbb{R})$ were
called Ogievetsky $n$, or Og $n$, generators in
\cite{fabiopeterogievetsky}.

Given the algebra of eqs. (\ref{KwithP}), (\ref{OgnwithP}) and
(\ref{OgnwithOgm}), we consider the group element written in the
form
  \begin{equation}
  g=e^{x^\m P_\m}\ldots
  e^{\Phi_{\m_1\ldots \m_n}^\n (x) K^{\m_1\ldots \m_n}{}_\n}\ldots e^{\Phi _{\m_1 \m_2}^\n (x)
  K^{\m_1 \m_2}{}_\n} e^{
  h_\m{}^\n (x) K^\m{}_\n}\quad , \label{groupelement}
  \end{equation}
where the momentum generator is contracted with the spacetime
coordinate $x^\m$, while all the other fields are functions of
$x^\m$. The fields $\Phi$ contracting the Og generators are called
Og fields. In particular, $\Phi_{\m_1 ...\m_{n+1}}^\n$ is an Og $n$
field.

We now consider the non-linear realisation of the algebra of eqs.
(\ref{KwithP}), (\ref{OgnwithP}) and (\ref{OgnwithOgm}) with as
local subalgebra the $D$-dimensional Lorentz algebra. We want the
theory to be invariant under transformations of $g$ of the form
  \begin{equation}
  g \rightarrow g_0 g h \quad , \label{gzerogh}
  \end{equation}
where $g_0$ is a constant group element and $h$ is a local Lorentz
group transformation. The fact that the group element transforms
under the Lorentz group from the right means that in the exponential
of $h_\m{}^\n$ we have to replace the column index with a Lorentz
index. As it will appear natural from the Maurer-Cartan form, we
identify the exponential of $h_\m{}^\n$ with the vierbein,
  \begin{equation}
  e_\mu{}^a= (e^h)_\mu{}^a \quad , \label{vierbein}
  \end{equation}
where the $a$ ($a=1,...,D$) index is a Lorentz index. This means
that the vierbein converts curved, that is $GL(D, \mathbb{R})$,
indices to flat, that is Lorentz, indices. We want local Lorentz
transformations, that act from the right on the group element, to
only rotate the vierbein, and it is for this reason that we have
written the group element with $h_\m{}^\n$ sitting on the far right.
One can show that acting as in eq. (\ref{gzerogh}) on $g$ one
reproduces general coordinate transformations for all the Og fields,
while the vierbein transforms under general coordinate
transformations and under local Lorentz transformations in the usual
way \cite{pashnev}.  This notation differs from the one used in
\cite{fabiopeterogievetsky}, where $GL(D,\mathbb{R})$ indices were
denoted with Latin letters.

The Maurer-Cartan form $g^{-1}d g$ is invariant under $g_0$
transformations in eq. (\ref{gzerogh}) and only transforms under
$h$. As a consequence, the generators have to be decomposed in
irreducible representations of the Lorentz algebra, and thus the
indices of the generators must be converted to Lorentz indices. One
gets
  \begin{equation}
  g^{-1}d g =  dx^\mu ( e_\mu{}^a P_a+ G_{\mu , a}{}^b K^a{}_b +G_{\mu ,a b}{}^c
  K^{ab}{}_c+\ldots ) \quad ,
  \end{equation}
with
  \begin{equation}
  G_{\mu , a}{}^b = (e^{-1}\partial_\mu
  e)_a{}^b -\Phi _{\mu \n}^\r (e^{-1})_a{}^\n
  e_\r{}^b   \label{christoffel}
  \end{equation}
  and
  \begin{equation}
  G_{\mu , a b}{}^c= (\partial_\mu \Phi _{\rho\kappa}^\lambda -2
  \Phi_{\mu\rho\kappa}^\lambda - \Phi_{\mu
  (\rho}^\tau \Phi_{\kappa ) \tau}^\lambda + {1 \over 2} \Phi_{\rho \kappa}^\tau
  \Phi_{\mu \tau}^\lambda ) (e^{-1})_a{}^\rho (e^{-1})_b{}^\kappa
  e_\lambda{}^c \quad .\label{riemannafterhiggs}
  \end{equation}
Lorentz indices can be raised and lowered using the invariant metric
$\eta_{ab}$. Moreover, apart for the momentum operator, the
generators belong to reducible Lorentz representations. In
particular the operator $K^{ab}$ splits in its antisymmetric part,
its symmetric traceless part and its trace, and the antisymmetric
part of $K^{ab}$ is the adjoint representation of the Lorentz
algebra. Note that nothing has happened to the generators as such.
The generators are invariant tensors, which one can think of as
constant matrices, and we have relabelled the indices of these
matrices according to the fact that we have to think about them as
invariant tensors of the Lorentz algebra.

Identifying as we anticipated in eq. (\ref{vierbein}) the vierbein
with the exponential of $h_\m{}^\n$, one realises that the quantity
$G_\m{}^{ab}$ defined in eq. (\ref{christoffel}) is part of the
covariant derivative of the vierbein if one further identifies
$\Phi_{\m\n}^\r$ with the Christoffel connection. In particular, if
one imposes that the symmetric part in $ab$ of $G_\m{}^{ab}$
vanishes, this forces to identify $\Phi_{\m\n}^\r$ with the
Levi-Civita connection \cite{kirsch}
  \begin{equation}
  \Phi _{\mu \nu}^\r= \Gamma_{\mu \nu}^\r \equiv {1\over
  2}g^{\r \tau}(\partial_\nu g_{\tau\mu}+
  \partial_\mu g_{\tau\nu}-\partial_\tau g_{\mu\nu}) \quad ,
  \label{levicivitaconnection}
  \end{equation}
and $G_\m{}^{ab}$ becomes the spin connection $\omega_\m{}^{ab} $
\cite{fabiopeterogievetsky},
  \begin{equation}
  \omega_{\mu}{}^{ab} ={1\over
  2} e^{\n a}(\partial_\mu e_\n{}^b -\partial_\n e_\mu{} ^b )
  - {1\over 2} e^{\n b }(\partial_\mu e_\n{}^a
  -\partial_\n e_\mu{}^a ) -{1\over 2} e^{ \n a } e^{ \r  b}
  (\partial_\n e_\r{}^c -\partial_\r
  e_\n{}^c )e_\mu{}^c  \quad , \label{spinconnection}
  \end{equation}
where we have denoted the inverse vierbein as
  \begin{equation}
  e^\m{}_a  = (e^{-1})_a{}^\m \quad .
  \end{equation}

In the term contracting the Og 1 generator, that is eq.
(\ref{riemannafterhiggs}), one can covariantly solve for the Og 2
field $\phi_{\m\n\r}^\lambda$ in terms of the Og 1 field, which is
the Levi-Civita connection in such a way that eq.
(\ref{riemannafterhiggs}) becomes the Riemann tensor
  \begin{equation}
  2G_{\mu,\rho\kappa}{}^\lambda= R_{\mu\rho}{}^\lambda{}_\kappa\equiv
  \partial_\mu \Gamma _{\rho\kappa}{}^\lambda-\partial_\rho
  \Gamma_{\mu\kappa}{}^\lambda +\Gamma_{\mu\tau}{}^\lambda \Gamma_{\rho
  \kappa}{}^\tau - \Gamma_{\rho\tau}{}^\lambda \Gamma_{\mu
  \kappa}{}^\tau
  \quad .
  \end{equation}
One can solve for the Og fields of any grade in terms of the lower
grade fields, which results in the Maurer-Cartan form only
containing the Riemann tensor and covariant derivatives thereof.
This concludes the review of section 2 of
\cite{fabiopeterogievetsky}.

The algebra of eqs. (\ref{KwithP}), (\ref{OgnwithP}) and
(\ref{OgnwithOgm}) can be deformed compatibly with $GL(D,
\mathbb{R})$. In particular, restricting our attention to the
generators up to Og 2, we can write the relevant commutators as
  \begin{eqnarray}
  & & [ K^\m{}_\n , P_\r ] = - \delta^\m_\r P_\n + a \delta^\m_\n P_\r
  \nonumber \\
  & & [ K^{\m_1 \m_2 }{}_\n , P_\r ] = \delta^{(\m_1}_\r K^{\m_2
  )}{}_\n + b \delta^{(\m_1}_\n K^{\m_2
  )}{}_\r + c \delta^{(\m_1}_\n \delta^{\m_2 )}_\r K \quad ,
  \end{eqnarray}
where $K$ is the trace of the $GL(D, \mathbb{R})$ generators,
  \begin{equation}
  K = K^\m{}_\m \quad .
  \end{equation}
The parameters $a$, $b$ and $c$ satisfy conditions coming from the
Jacobi identities. In particular, if $Da \neq 1$, one can without
loss of generality impose $b=0$, and then determine $c$ to be
  \begin{equation}
  c = \frac{a}{1 -Da} \quad .
  \end{equation}
Summarising, the deformed algebra is
  \begin{eqnarray}
  & & [ K^\m{}_\n , P_\r ] = - \delta^\m_\r P_\n + a \delta^\m_\n P_\r
  \nonumber \\
  & & [ K^{\m_1 \m_2 }{}_\n , P_\r ] = \delta^{(\m_1}_\r K^{\m_2
  )}{}_\n  + \frac{a}{1-Da} \delta^{(\m_1}_\n \delta^{\m_2 )}_\r K
  \label{modifiedgravity}
  \end{eqnarray}
for any parameter $a$, provided that $Da \neq 1$.

We now consider the group element of eq. (\ref{groupelement}), and
we compute the Maurer-Cartan form using the modified commutators of
eq. (\ref{modifiedgravity}). The result is
  \begin{equation}
  g^{-1}d g =  dx^\mu ( e^{-ah} ( e^h )_\mu{}^a P_a+ G_{\mu , a}{}^b K^a{}_b +\ldots ) \quad ,
  \end{equation}
where we have defined
  \begin{equation}
  h = h_\m{}^\m \quad ,
  \end{equation}
and where
  \begin{equation}
  G_{\mu , a}{}^b = (e^{-h}\partial_\mu
  e^h )_a{}^b -\Phi _{\mu \n}^\r (e^{-h})_a{}^\n
  ( e^h )_\r{}^b  -\frac{a}{1 -Da} \Phi_{\m \n}^\n \delta_a^b \quad . \label{christoffelmodified}
  \end{equation}
We now interpret the matrix contracting the momentum operator as the
vierbein,
  \begin{equation}
  e_\mu{}^a= e^{-ah} ( e^h )_\mu{}^a \quad ,
  \label{modifiedvierbeinwitha}
  \end{equation}
and inverting this relation one gets
  \begin{equation}
  ( e^h )_\mu{}^a  = ( {\rm det} e )^{\frac{a}{1-Da}} e_\mu{}^a \quad
  ,
  \end{equation}
where we have denoted the determinant of the vierbein with ${\rm det
}e$ to avoid confusion as much as possible between Euler's number
and the vierbein. If we plug this relation into eq.
(\ref{christoffelmodified}), we get
  \begin{equation}
  G_{\mu , a}{}^b = (e^{-1}\partial_\mu
  e )_a{}^b + \frac{a}{1 -Da} ( {\rm det} e )^{-1} \de_\m   ( {\rm det}
  e )
  \delta_a^b
   -\Phi _{\mu \n}^\r (e^{-1})_a{}^\n
  ( e )_\r{}^b  -\frac{a}{1 -Da} \Phi_{\m \n}^\n \delta_a^b \quad
  .\label{modifiedwithaandchristoffel}
  \end{equation}
If we now impose that the symmetric part in $ab$ of this equation
vanishes, we find that eq. (\ref{levicivitaconnection}) is still a
solution, and the $\delta_a^b$ part of eq.
(\ref{modifiedwithaandchristoffel}) cancels because eq.
(\ref{levicivitaconnection}) gives the well-known formula
   \begin{equation}
   \Phi_{\m \n}^\n = \Gamma^\n_{\m\n} = ( {\rm det} e )^{-1} \de_\m   ( {\rm det}
  e ) \quad .
  \end{equation}
This proves that the modification of the algebra of diffeomorphisms
as in eq. (\ref{modifiedgravity}) is equivalent to the redefinition
of the vierbein in terms of $h_\m{}^\n$ as in eq.
(\ref{modifiedvierbeinwitha}).

Before we conclude this section, we want to make a comment on
dimensional reduction. We consider a circle dimensional reduction
from dimension $D+1$ to dimension $D$, we denote with $\m$ and $a$
the curved and flat indices in $D$ dimensions, and we denote with
$y$ the $D+1$-th coordinate. The $D+1$ dimensional momentum splits
in $P_\m$ and $Q= P_y$. As shown in \cite{fabiopeterogievetsky},
circle dimensional reduction corresponds to a truncation of the
algebra in which the operator $Q$ is projected out, and consistently
one must project out all the generators that have non-trivial
commutator with $Q$. By looking at eq. (\ref{KwithP}), this implies
that $K^y{}_\m$ must be projected out. This implies the standard
ansatz for the $D$ dimensional vierbein,
  \begin{equation}
  \left( \begin{array}{cc}
  e^{\a \phi} e_\mu{}^a  & e^{\b \phi} A_\mu   \\
  0 & e^{\b \phi} \end{array} \right) \quad ,\label{vierbeinansatz}
  \end{equation}
and computing the part of the Maurer-Cartan form along $d x^\m$,
neglecting for simplicity the Og contribution, one gets
  \begin{equation}
  d x^\m g^{-1} \de_\m g = e^{\a \phi} e_\m{}^a P_a + ( e^\n{}_a
  \de_\m e_\n{}^b + \a \delta_a^b \de_\m \phi ) K^a{}_b + e^{(\b-\a
  )\phi} \de_\m A_\n e^\n{}_a K^a{}_y + \b \de_\m \phi K^y{}_y
  \quad . \label{gravityreducedanyframe}
  \end{equation}
By looking at this equation, we define the $D$ dimensional vector
and scalar generators as
  \begin{eqnarray}
  & & R^\m =  K^\m{}_y \nonumber \\
  & & R = \a K + \b K^y{}_y \quad ,
  \end{eqnarray}
and the non-trivial commutators, apart from the commutators with the
$GL(D, \mathbb{R})$ generators which are standard, are
  \begin{eqnarray}
  & & [ R , R^\m ] = (\a -\b ) R^\m \nonumber \\
  & & [ R , P_\m ] = -\a P_\m \quad .
  \end{eqnarray}
From these commutators it is then easy to show that the
Maurer-Cartan form of eq. (\ref{gravityreducedanyframe}) arises from
the $D$ dimensional group element
  \begin{equation}
  g = e^{x \cdot P} e^{A_\m R^\m } e^{h_\m{}^\n K^\m{}_\n} e^{\phi
  R} \quad .
  \end{equation}
This concludes the analysis of this section. In the next section we
will consider a (generalised) dimensional reduction of eleven
dimensional supergravity to ten dimensions, and for simplicity we
will work in the frame in which $\a = 0 $ and $\b =1$, but all the
results can easily be generalised to any frame.

\section{Local $E_{11}$ and gauged IIA}
In \cite{fabiopeterogievetsky} it was shown that the Scherk-Schwarz
reduction of the IIB theory corresponds to a non-linear realisation
based on an $E_{11}$ group element that is entirely
nine-dimensional, apart from an overall transformation with respect
to the generators of the internal symmetry of the IIB theory which
is linear in the compactified coordinate. The main aim of this
section is to perform for the gauged IIA theory of \cite{hlw} an
analysis equivalent to the one performed in
\cite{fabiopeterogievetsky} for the Scherk-Schwarz reduction of the
IIB theory. This analysis is motivated by ref. \cite{fibrebundles},
where the gauged IIA theory was derived performing a generalised
Scherk-Schwarz reduction of eleven dimensional supergravity in which
one performs a scaling transformation of the fields which is linear
in the internal coordinate. The symmetry under rescaling of the
fields was called ``trombone'' symmetry in \cite{fibrebundles},
because although it is a symmetry of the field equations, it
actually gives rise to an overall scaling of the lagrangian. The
fact that this symmetry is not a symmetry of the lagrangian implies
that its gauging results in a theory which does not admit a
lagrangian formulation.

We first review the $E_{11}$ analysis of 11-dimensional supergravity
with the inclusion of the Og generators, as was derived originally
in \cite{fabiopeterogievetsky}. We use a notation similar to the one
of the previous section, and we thus use Greek letters to denote the
$GL(D , \mathbb{R})$ indices. This notation again differs from the
one used in \cite{fabiopeterogievetsky}. In particular,
$GL(11,\mathbb{R})$ indices are denoted by $\hat{\m}$ ($\hat{\m} = 1
, ...,11$), and similarly 11-dimensional Lorentz indices are denoted
by $\hat{a}$. We only consider the $GL(11, \mathbb{R})$ generator
$K^{\hat{\m}}{}_{\hat{\n}}$ and the 3-form generator $R^{\hat{\m}_1
\hat{\m}_2 \hat{\m}_3}$, which corresponds to a truncation of the
$E_{11}$ algebra to level 1 (and only considering positive level
generators). The relevant $E_{11}$ commutators are thus the
commutators giving the $GL(11,\mathbb{R})$ algebra and
  \begin{equation}
  [ K^{\hat{\m}}{}_{\hat{\n}} , R^{\hat{\r}_1 \hat{\r}_2 \hat{\r}_3} ]
  = 3 \delta^{[\hat{\r}_1}_{\hat{\n}} R^{ | \hat{\m} | \hat{\r}_2
  \hat{\r}_3 ]} \quad .
  \end{equation}
As explained in \cite{fabiopeterogievetsky}, in order to promote the
3-form constant shift to a gauge transformation, we have to add an
infinite set of Og generators, the first one being $K_1^{\hat{\m},
\hat{\n}_1 \hat{\n}_2 \hat{\n}_3}$ satisfying
\begin{equation}
K_1^{\hat{\m}, \hat{\n}_1 \hat{\n}_2 \hat{\n}_3} = K_1^{\hat{\m},[
\hat{\n}_1 \hat{\n}_2 \hat{\n}_3 ]} \qquad K_1^{[\hat{\m} ,
\hat{\n}_1 \hat{\n}_2 \hat{\n}_3
   ]}=0 \quad ,
\end{equation}
whose commutator with momentum is
\begin{equation}
[ K_1^{\hat{\m} , \hat{\n}_1 \hat{\n}_2 \hat{\n}_3} , P_{\hat{\r} }]
= \delta^{\hat{\m}}_{\hat{\r}} R^{\hat{\n}_1 \hat{\n}_2 \hat{\n}_3}
- \delta^{ [ \hat{\m} }_{\hat{\r}}
   R^{\hat{\n}_1 \hat{\n}_2 \hat{\n}_3 ]} \quad .
\end{equation}
If one considers the group element in the form
\begin{equation}
g = e^{x \cdot P} e^{\Phi_{\rm Og} K^{\rm Og}}
e^{A_{\hat{\m}\hat{\n}\hat{\r}} R^{\hat{\m}\hat{\n}\hat{\r}}}
e^{h_{\hat{\m}}{}^{\hat{\n}} K^{\hat{\m}}{}_{\hat{\n}}}
\end{equation}
and computes the Maurer-Cartan form, one gets
\begin{equation}
g^{-1} d g = d x^{\hat{\m }}\left[ e_{\hat{\m}}{}^{\hat{a}}
P_{\hat{a}} +\left( \de_{\hat{\m}} A_{\hat{\n}\hat{\r}\hat{\s}} -
\Phi_{\hat{\m} , \hat{\n}\hat{\r}\hat{\s}} \right)
e^{\hat{\n}}{}_{\hat{a}} e^{\hat{\r}}{}_{\hat{b}}
e^{\hat{\s}}{}_{\hat{c}} R^{\hat{a}\hat{b}\hat{c}} + ... \right]
\end{equation}
where the dots denote both the Og generators contribution, as well
as the gravity sector which is as in \cite{fabiopeterogievetsky} and
reviewed in section 2. As explained in \cite{fabiopeterogievetsky},
the inverse Higgs mechanism permits to solve covariantly for the Og
1 field $\Phi_{\hat{\m} , \hat{\n}\hat{\r}\hat{\s}} $ in terms of
the derivative of the 3-form potential, in such a way that only the
completely antisymmetric term $\de_{[\hat{\m}}
A_{\hat{\n}\hat{\r}\hat{\s}]}$ survives, which is the field strength
of the 3-form. Similarly, by the same mechanism the Og generators
are contracted with covariant derivatives of the field strength of
the  3-form.

We now consider the dimensional reduction of this system to ten
dimensions. For simplicity we take $\alpha=0 $ and $ \beta=1 $ in
the vierbein ansatz of eq. (\ref{vierbeinansatz}). The construction
is easy to generalise to any other frame. The notation for the
dimensionally reduced gravity generators is the same as in the
previous section, while the 11-dimensional 3-form generator gives
rise to the 3-form $R^{\m_1 \m_2 \m_3 }$ and the 2-form $R^{\m_1
\m_2 } = R^{\m_1 \m_2 y}$, where the Greek index $\m$ is a
$GL(10,\mathbb{R})$ index and $y$ denotes the 11-th direction.

In terms of these generators, the $E_{11}$ algebra becomes (we only
consider the non-vanishing commutators)
\begin{equation}
[ K^\m{}_\n , R^\r ] = \delta^\r_\n R^\m \qquad [ K^\m{}_\n ,
R^{\r_1 \r_2} ] = 2 \delta^{[\r_1}_\n R^{|\m |\r_2 ]} \qquad  [
K^\m{}_\n , R^{\r_1 \r_2 \r_3} ] = 3 \delta^{[\r_1}_\n R^{|\m |\r_2
\r_3 ]}
\end{equation}
\begin{equation}
[ R, R^\m ] = - R^\m \qquad \qquad \quad  \ \ \  [ R , R^{\m\n} ] =
R^{\m\n}
\end{equation}
\begin{equation}
[ R^\m , R^{\n\r} ]= R^{\m\n\r} \quad .
\end{equation}
We also have
  \begin{equation}
  [R^\m , P_\n ] = -\delta^\m_\n Q \quad ,
  \end{equation}
where as in the previous section $Q$ denotes the momentum operator
in the $y$ direction.

The 11-dimensional Og generator $K^{\hat{\m} , \hat{\n}_1 \hat{\n}_2
\hat{\n}_3 }$ gives rise to the 10-dimensional Og generators $K^{\m,
\n_1 \n_2 \n_3}$, $K^{\m ,\n_1 \n_2}$, $K^{\m_1 \m_2 \m_3}$ and
$K^{\m_1 \m_2}$, with
\begin{eqnarray}
& & K^{\m , \n_1 \n_2} = K^{\m , \n_1 \n_2 y} -  K^{[\m , \n_1 \n_2
]y} \nonumber \\
& & K^{\m_1 \m_2 \m_3} = \frac{4}{3} K^{y , \m_1 \m_2 \m_3} \nonumber \\
& & K^{\m_1 \m_2} = K^{y , \m_1 \m_2 y} \quad .
\end{eqnarray}
The commutators of these operators with $P_\m$ and $Q$ are
\begin{eqnarray}
& & [ K^{\m , \n_1 \n_2 \n_3} , P_\r ]=  \delta^\m_\r R^{\n_1 \n_2
\n_3} - \delta^{[ \m}_\r R^{\n_1 \n_2 \n_3 ]} \qquad \quad [ K^{\m ,
\n_1 \n_2 \n_3}, Q] = 0 \nonumber \\
& & [ K^{\m_1 \m_2 \m_3 } , P_\n ] = \delta^{[\m_1}_\n R^{\m_2 \m_3
]} \quad \qquad \qquad \qquad \quad \ \  [ K^{\m_1 \m_2 \m_3 }, Q ]
= R^{\m_1 \m_2 \m_3 }
\nonumber \\
& & [ K^{\m , \n_1 \n_2 } , P_\r ] = \delta^\m_\r R^{\n_1 \n_2 }
-\delta^{ [ \m}_\r R^{\n_1 \n_2 ]}  \quad \quad \qquad \  \ \ [
K^{\m , \n_1 \n_2
} , Q ] =0 \nonumber \\
& & [K^{\m_1 \m_2 } , P_\n ] = 0 \quad \qquad \qquad \qquad \quad
\qquad \ \ \qquad \  [K^{\m_1 \m_2 } , Q ] = R^{\m_1 \m_2 } \quad .
\end{eqnarray}
We also consider the 10-dimensional Og generators that arise from
the 11-dimensional gravity Og 1 generator $K^{\m \n}{}_\r$. In
particular we are only interested in the Og generators whose lower
index in is the $y$ direction, that are
  \begin{equation}
  K^{\m\n} = K^{\m\n}{}_y \quad \quad K^{\m} = 2 K^{\m y}{}_y \quad \quad K =
  K^{yy}{}_y \quad ,
  \end{equation}
and whose commutation relation with $P_\m$ and $Q$ are
  \begin{eqnarray}
  & & [ K^{\m\n} , P_\r ] = \delta^{(\m}_\r R^{\n )} \quad \quad \  [
  K^{\m\n} , Q ] =0 \nonumber \\
  & & [ K^\m , P_\n ] = \delta^\m_\n R \quad \quad \qquad \, [ K^\m , Q ] =
  R^\m
  \nonumber \\
  & & [ K , P_\m ] = 0 \qquad \qquad \qquad [ K , Q ] = R \quad .
  \end{eqnarray}

We now consider the non-linear realisation based on this algebra. We
first consider the case of standard massless dimensional reduction,
which corresponds to taking the group element
  \begin{equation}
  g  = e^{x\cdot P} e^{y Q} e^{\Phi_{\rm Og} K^{\rm Og}} e^{A_{\m\n\r}
  R^{\m\n\r} } e^{A_{\m\n} R^{\m\n}} e^{A_\m R^\m} e^{\phi R}
  e^{h_\m{}^\n K^\m{}_\n} \label{masslessIIAgroupelement} \quad ,
  \end{equation}
where we take all the fields not to depend on $y$. We then compute
the Maurer-Cartan form
\begin{equation}
g^{-1} d g = d x^\m g^{-1} \de_\m g + d y g^{-1} \de_y g \quad .
\end{equation}
We first consider the part along $dx^\m$. Following
\cite{fabiopeterogievetsky}, we use the inverse Higgs mechanism to
covariantly solve for the not fully antisymmetric Og fields in terms
of the other fields in such a way that all the terms in the
Maurer-Cartan form are completely antisymmetric. This gives
  \begin{eqnarray}
  & & dx^\m g^{-1} \de_\m g = d x^\m [ e_\m{}^a P_a + e^\phi A_\m Q
  + \left( \de_\m \phi - \Phi_\mu \right) R + e^\phi \de_{[\m} A_{\n
  ]} e^\n{}_a R^a \nonumber \\
  & & \quad   + e^{-\phi } \left( \de_{[\m} A_{\n\r ]} -
  \Phi_{\m\n\r} - \Phi_{[\m} A_{\n\r ]} \right) e^\n{}_a e^\r{}_b
  R^{ab}
  \nonumber \\
  & & \quad  + \left( \de_{[\m} A_{\n\r\s ]} - \de_{[\m} A_{\n\r}A_{\s
  ]} +\Phi_{[\m \n\r} A_{\s ]} + \Phi_{[\m} A_{\n\r} A_{\s ]}
  \right) e^\n{}_a e^\r{}_b e^\s{}_c R^{abc} + ... ]
  \label{dxmupart}
  \end{eqnarray}
We then consider the $dy$ term. Again following
\cite{fabiopeterogievetsky} we impose that the part of the
Maurer-Cartan form in the $dy$ direction vanishes apart from the $Q$
term. This imposes that all the Og fields associated to the Og
generators that do not commute with $Q$ must vanish:
  \begin{equation}
  \Phi = 0 \quad \Phi_\m = 0 \quad \Phi_{\m_1 \m_2 } = 0 \quad
  \Phi_{\m_1 \m_2 \m_3} =0 \quad .
  \end{equation}
Plugging these conditions into eq. (\ref{dxmupart}), we then read
the field strengths
  \begin{eqnarray}
  & & F_{\m\n} = \de_{[\m} A_{\n
  ]} \nonumber \\
  & & F_{\m\n\r} =  \de_{[\m} A_{\n\r ]} \nonumber \\
  & & F_{\m\n\r\s} = \de_{[\m} A_{\n\r\s ]} - \de_{[\m} A_{\n\r}A_{\s
  ]}\quad , \end{eqnarray}
that are the field strengths of the gauge fields of the massless IIA
theory. Acting with $g_0$ transformations on the group element of
eq. (\ref{masslessIIAgroupelement}) one also derives the
corresponding gauge transformations, that are
  \begin{eqnarray}
  & & \delta A_{\m} = \de_\m \Lambda \nonumber \\
  & & \delta A_{\m\n} = \de_{[\m} \Lambda_{\n ]} \nonumber \\
  & & \delta A_{\m\n\r} = \de_{[\m} \Lambda_{\n\r ]} + \de_{[\m}
  \Lambda A_{\n\r ]}\quad .
  \end{eqnarray}

We now want to derive the field strengths and gauge transformations
of the gauged IIA theory of \cite{hlw,fibrebundles} in an analogous
way. We take as our starting point an 11-dimensional group element
that has a non-trivial $y$ dependence, namely
  \begin{equation}
  g  = e^{x\cdot P} e^{y Q} e^{m y (K + R)} e^{\Phi_{\rm Og} K^{\rm Og}} e^{A_{\m\n\r}
  R^{\m\n\r} } e^{A_{\m\n} R^{\m\n}} e^{A_\m R^\m} e^{\phi R}
  e^{h_\m{}^\n K^\m{}_\n} \quad ,\label{tromboneIIAgroupelement}
  \end{equation}
where $K$ is the trace of the $GL(10,\mathbb{R})$ generators, $m$ is
a constant parameter and we take all the fields not to depend on
$y$. Observe that this particular choice of the group element is due
to the fact the trombone scaling is generated by
$K^{\hat{\m}}{}_{\hat{\m}}$ in eleven dimensions, and
$K^{\hat{\m}}{}_{\hat{\m}} = K + R$ in the frame in which $\a=0$ and
$\beta =1$ in eq. (\ref{vierbeinansatz}). One can easily generalise
this to an arbitrary frame.

We now compute the Maurer-Cartan form. As in the massless case, we
first consider the $dx^\m$ term, and we use the inverse Higgs
mechanism to solve for the Og fields with mixed symmetry in terms of
the other fields in such a way that all the terms that are left in
the Maurer-Cartan form are completely antisymmetric. With respect to
eq. (\ref{dxmupart}) the $P_\m$ and $Q$ terms, as well as the Og
fields, acquire a $y$ dependence due to the non-trivial form of the
group element of eq. (\ref{tromboneIIAgroupelement}). The result is
  \begin{eqnarray}
  & & dx^\m g^{-1} \de_\m g = d x^\m [ e^{my} e_\m{}^a P_a + e^{my} e^\phi A_\m Q
  + \left( \de_\m \phi - e^{my}\Phi_\mu \right) R + e^\phi \de_{[\m} A_{\n
  ]} e^\n{}_a R^a \nonumber \\
  & & \  + e^{-\phi } \left( \de_{[\m} A_{\n\r ]} -
  e^{my} \Phi_{\m\n\r} - e^{my} \Phi_{[\m} A_{\n\r ]} \right) e^\n{}_a e^\r{}_b
  R^{ab}
  \nonumber \\
  & & \  + \left( \de_{[\m} A_{\n\r\s ]} - \de_{[\m} A_{\n\r}A_{\s
  ]} +\e^{my} \Phi_{[\m \n\r} A_{\s ]} + e^{my} \Phi_{[\m} A_{\n\r} A_{\s ]}
  \right) e^\n{}_a e^\r{}_b e^\s{}_c R^{abc} + ... ]
  \label{dxmuparttrombone}
  \end{eqnarray}
The fact that this term has a non-trivial $y$ dependence is the
crucial difference with respect to the Scherk-Schwarz reduction of
IIB discussed in \cite{fabiopeterogievetsky}. In that case, the
group element was deformed by a $y$-dependent $SL(2 ,\mathbb{R})$
transformation, which commutes with momentum. Correspondingly, the
$dx^\m$ part of the Maurer-Cartan form did not contain any $y$
dependence. This is what guarantees the consistency of the
truncation to the lower dimensional theory. In this case, the
$dx^\m$ part of the Maurer-Cartan contains a $y$ dependence, and
this is the translation in this group-theoretic language of the fact
that the trombone symmetry is not a symmetry of the lagrangian but
only of the field equations. As emphasised in \cite{fibrebundles},
having such a symmetry is actually sufficient to guarantee that also
in this case the truncation to ten dimensions is consistent at the
level of the field equations. We will see in the following how this
notion of  consistency of the truncation is translated in our
langauge.

We now compute the $dy$ part of the Maurer-Cartan form. We get
  \begin{eqnarray}
  & & dy g^{-1} \de_y g = e^\phi e^{my} Q + m (K + R ) -
  e^{my} \Phi R  +  e^\phi e^{my} \left( - \Phi_\m + \Phi
  A_\mu \right) e^\m{}_a R^a \nonumber \\
  & & \ \ + e^{-\phi} \left( -e^{my} \Phi
  A_{\m\n} - e^{my} \Phi_{\m\n} + 3 m A_{\m\n} \right) e^\m{}_a
  e^\n{}_b R^{ab} \nonumber\\
  & & \ \ + ( - e^{my} \Phi_\m A_{\n\r} + e^{my} \Phi
  A_{\m\n} A_{\r} - e^{my} \Phi_{\m\n\r} + e^{my} \Phi_{\m\n} A_{\r}
  + 3m A_{\m\n\r} \nonumber \\
  & & \ \ - 3 m A_{\m\n} A_{\r} ) e^\m{}_a e^\n{}_b
  e^\r{}_b R^{abc} \quad . \label{dytrombone}
  \end{eqnarray}
Following \cite{fabiopeterogievetsky}, we now use the inverse Higgs
mechanism to impose that all the terms in eq. (\ref{dytrombone})
proportional to positive level generators vanish. This gives
  \begin{equation}
  \Phi = 0 \quad \qquad \Phi_\m =0 \quad , \label{sameasbefore}
  \end{equation}
as well as
  \begin{equation}
  e^{my} \Phi_{\m\n}  -3 m A_{\m\n} =0 \qquad \quad  e^{my} \Phi_{\m\n\r}  -3 m A_{\m\n\r} =0
  \quad . \label{ogpropfieldmass}
  \end{equation}
Substituting these relations in eq. (\ref{dxmuparttrombone}), we
then read the field strengths
  \begin{eqnarray}
  & & F_{\m\n} = \de_{[\m} A_{\n
  ]} \nonumber \\
  & & F_{\m\n\r} =  \de_{[\m} A_{\n\r ]} - 3 m A_{\m\n\r} \nonumber \\
  & & F_{\m\n\r\s} = \de_{[\m} A_{\n\r\s ]} - \de_{[\m} A_{\n\r}A_{\s
  ]} + 3 m A_{[\m\n\r} A_{\s ]} \quad ,
  \label{fieldstrengthtrombone}
  \end{eqnarray}
that are the field strengths of the gauge fields of the gauged IIA
theory \cite{hlw,fibrebundles}. Acting with $g_0$ transformations on
the group element of eq. (\ref{masslessIIAgroupelement}) one also
derives the gauge transformations
  \begin{eqnarray}
  & & A_{\m} \rightarrow A_\m + \de_\m \Lambda \nonumber \\
  & & A_{\m\n} \rightarrow e^{3m\Lambda} A_{\m\n} +  \de_{[\m} \Lambda_{\n ]} \nonumber \\
  & & A_{\m\n\r} \rightarrow e^{3m\Lambda} A_{\m\n\r} + \de_{[\m} \Lambda_{\n\r ]} + \de_{[\m}
  \Lambda A_{\n\r ]}\quad ,
  \end{eqnarray}
which transform covariantly the field strengths of eq.
(\ref{fieldstrengthtrombone}), that is
  \begin{eqnarray}
  & & F_{\m\n} \rightarrow F_{\m\n} \nonumber \\
  & & F_{\m\n\r} \rightarrow e^{3m\Lambda} F_{\m\n\r} \nonumber \\
  & & F_{\m\n\r\s} \rightarrow e^{3m\Lambda} F_{\m\n\r\s} \quad .
  \end{eqnarray}

We now perform an analysis of the deformed algebra that parallels
the one performed in \cite{fabiopeterogievetsky} for the case of the
Scherk-Schwarz reduction of IIB to nine dimensions. We start
observing that eq. (\ref{ogpropfieldmass}) relates the Og fields to
the $E_{11}$ fields times the deformation parameter. Iterating this
one obtains for any $n$ an Og $n$ field identified with an $E_{11}$
field times the $n$th power of the mass parameter. This generalises
to all the fields in the theory whose corresponding operators have
non-vanishing commutator with the operator $K +R $. Putting these
solutions into the original group element of eq.
(\ref{tromboneIIAgroupelement}) we find that it takes the form
  \begin{equation}
  g  = e^{x \cdot P} e^{yQ} e^{y ( K+ R )}  e^{\Phi_{\rm Og} \tilde{K}^{\rm Og}}
  e^{A_{\m\n\r} \tilde{R}^{\m\n\r}} e^{A_{\m\n} \tilde{R}^{\m\n}}
  e^{A_\m  R^\m} e^{\phi R } e^{h_\m{}^\n K^\m{}_\n} \quad ,
  \end{equation}
where
  \begin{eqnarray}
  & & \tilde{R}^{\m\n} = {R}^{\m\n} + 3m e^{-my} K^{\m\n} + ...
  \nonumber \\
  & & \tilde{R}^{\m\n\r} = {R}^{\m\n\r} + 3 m e^{-my} K^{\m\n\r} +... \quad ,
  \end{eqnarray}
where the dots correspond to higher powers in $m$ multiplying higher
grade Og generators, and $\tilde{K}$ denotes deformed Og generators
associated with ten-dimensional gauge transformations. We also
define, as suggested by the Maurer-Cartan form of eq.
(\ref{dxmuparttrombone}), the deformed 10-dimensional momentum
operator as
  \begin{equation}
  \tilde{P}_\m = e^{my} P_\m \quad .
  \end{equation}
We therefore get the commutator
  \begin{equation}
  [ \tilde{R}^{\m_1 \m_2 \m_3 } , \tilde{P}_\n ] = 3m
  \delta^{[\m_1}_\n \tilde{R}^{\m_2 \m_3 ] } \quad ,
  \label{deformed3pisdeformed2}
  \end{equation}
while the commutator of $\tilde{R}^{\m_1 \m_2}$ with $\tilde{P}_\m$
vanishes.

We think of the deformed generators constructed this way as
constituting a deformed local $E_{11}$ algebra.  This deformed
algebra has an algebraic classification as the set of generators
that commute with the operator
  \begin{equation}
  \tilde{Q} = e^{my}Q + m (K+ R )
  \quad .
  \end{equation}
This operator can be read from eq. (\ref{dytrombone}), which indeed
becomes, once one imposes the conditions of eqs.
(\ref{sameasbefore}) and (\ref{ogpropfieldmass}),
  \begin{equation}
  d y e^{-\phi R} \tilde{Q} e^{\phi R} \quad .
  \end{equation}
In terms of the operator $\tilde{Q}$ the commutator between $R^\m$
and $\tilde{P}_\m$ reads
 \begin{equation}
 [ R^\m , \tilde{P}_\n ] = - \delta^\m_\n \tilde{Q} + m \delta^\m_\n
 ( K+R) \quad . \label{RwithPisKplusR}
 \end{equation}

We then consider the scalar sector of eq. (\ref{dxmuparttrombone}),
that is
  \begin{equation}
  e^\phi e^{my} A_\m Q + \de_\m \phi R = A_\m e^{-\phi R} \tilde{Q}
  e^{\phi R} + ( \de_\m \phi - m A_\m ) R - m A_\m K \quad .
  \end{equation}
The $R$ term in this equation gives the covariant derivative for the
scalar,
  \begin{equation}
  D_\m \phi = \de_\m \phi - m A_\m \quad ,
  \end{equation}
which is invariant under
  \begin{equation}
  \delta \phi = m \Lambda \quad \qquad \delta A_\m = \de_\m \Lambda
  \quad .
  \end{equation}

Finally, we consider the gravity sector. This is again different
with respect to the Scherk-Schwarz reduction discussed in
\cite{fabiopeterogievetsky}. Indeed, in that case the deformation of
the group element was due to an internal symmetry generator, which
commutes with the gravity generators, and thus the analysis of the
dimensionally reduced gravity sector was trivial. In this case the
deformation involves the generator $K$, which is the trace of the
$GL(10,\mathbb{R})$ generators, and as such this has a non-trivial
effect in the gravity sector. Specifically, taking into account the
$K$ term in the Maurer-Cartan form, the $K^{ab}$ term becomes
  \begin{equation}
  \left[ (e^{-1} \de_\m e )_{ab} - \Phi_{\m\n}^\r e^\n_a e_{\r b} -
  m A_\m \eta_{ab} \right ] K^{ab} \quad . \label{gravitydeformed}
  \end{equation}
Observing that the vierbein transforms under $\Lambda$ as
  \begin{equation}
  e_\m{}^a \rightarrow e^{m\Lambda} e_{\m}{}^a \quad ,
  \label{vierbeintransf}
  \end{equation}
we write the term contracting $K^{ab}$ in eq.
(\ref{gravitydeformed}) as
  \begin{equation}
  (e^{-1} D_\m e )_{ab} - \Phi_{\m\n}^\r e^\n_a e_{\r b}
   \quad , \label{deformedspinconnectionchristoffel}
  \end{equation}
where $D_\m$ is the derivative covariantised with respect to the
transformation of eq. (\ref{vierbeintransf}), that is $D_\m = \de_\m
- m A_\m$. Applying the same arguments of
\cite{fabiopeterogievetsky}, which are reviewed in section 2, we
obtain that imposing that the symmetric part in $ab$ of eq.
(\ref{deformedspinconnectionchristoffel}) vanishes gives for the
antisymmetric part the spin connection as in eq.
(\ref{spinconnection}), but with the derivative $\de_\m$ substituted
by the covariant derivative $D_\m$. This is \cite{henning}
 \begin{equation}
 \tilde{\omega}_\m{}^{ab} = \omega_\m{}^{ab} - 2 m e_\m{}^{[a} e^{| \n
 | b]} A_\n \quad . \label{deformedspinconnection}
 \end{equation}
If one plugs this into the Maurer-Cartan form and applies the
inverse Higgs mechanism at the level of the next gravity Og field,
one obtains that the term contracting $K^{ab}{}_c$ is the
covariantised Riemann tensor
  \begin{equation}
  \tilde{R}_{\m\n}{}^{ab} = 2 \de_{[\m} \tilde{\omega}_{\n ]}{}^{ab} + 2
  \tilde{\omega}_{[\m}^{ac} \tilde{\omega}_{\n ] c}{}^b \quad .
  \label{deformedriemanntensor}
  \end{equation}
Therefore this reproduces exactly the field theory analysis of
\cite{henning} in the gravity sector.

The question we now want to address is in what sense one can
truncate the algebra in such a way that the resulting theory is
purely ten-dimensional. What we want to do is to project out of the
algebra the operator $\tilde{Q}$, and consider the group element as
a purely ten-dimensional one with commutation relations deformed
with respect to the massless case. From eq. (\ref{RwithPisKplusR})
we consider as a starting point for the ten-dimensional deformed
algebra the commutator
  \begin{equation}
  [ R^{\m} , P_\n ] = m \delta^\m_\n (K +R ) \quad ,
  \label{deformedinten}
  \end{equation}
where now we have for simplicity dropped the tilde from the deformed
generators. We want to determine the rest of the algebra by
requiring the closure of the Jacobi identities. This is exactly the
method explained in \cite{fabiopeterogievetsky} and applied in
\cite{hierarchyE11} to derive the deformed algebra associated to any
gauged maximal supergravity in any dimension.

The Jacobi identity involving $K+R$, $R^\m$ and $P_\m$ gives
  \begin{equation}
  [ R , P_\m ] = P_\m \quad ,
  \end{equation}
which implies
  \begin{equation}
  [ K+R , P_\m ] =0 \quad .
  \end{equation}
We then get
  \begin{equation}
  [ R^{\m\n} , P_\r ] = 0 \quad \qquad [ R^{\m\n\r} , P_\s ] = 3 m
  \delta^{[\m}_\s R^{\n\r ]} \quad .
  \end{equation}
We thus recover the commutation relation of eq.
(\ref{deformed3pisdeformed2}) from a purely ten-dimensional
perspective. If we then consider the ten-dimensional group element
  \begin{equation}
  g = e^{ x \cdot P} e^{\Phi_{\rm Og} K^{\rm Og} }
  e^{A_{\m\n\r}R^{\m\n\r}} e^{A_{\m\n} R^{\m\n}} e^{A_\m R^\m}
  e^{\phi R} e^{h_\m{}^\n K^\m{}_\n } \quad ,
  \end{equation}
the corresponding Maurer-Cartan form gives, once the inverse Higgs
mechanism is applied, the field strengths of eq.
(\ref{fieldstrengthtrombone}), as well as the covariantised spin
connection of eq. (\ref{deformedspinconnection}) and the
covariantised Riemann tensor of eq. (\ref{deformedriemanntensor}).

As a final comment, we discuss the overlap of this deformation,
corresponding to the gauged IIA theory, with the deformation
associated with Romans massive IIA theory. Denoting with $m_R$ the
mass parameter associated to Romans theory, the deformation
corresponds to a non-vanishing commutator between the 2-from
generator and momentum \cite{igorpeterromans}
  \begin{equation}
  [ R^{\m\n} , P_\r ] = m_R \delta^{[ \m}_\r R^{\n ]} \quad .
  \end{equation}
A simple computation shows that using this commutator together with
the one of eq. (\ref{deformedinten}) the Jacobi identity involving
$R^{\m\n}$, $P_\r$ and $P_\s$ closes only if the quadratic
constraint
  \begin{equation}
  m m_R =0
  \end{equation}
holds. This means that it is not consistent to turn on both
deformations together. This result is perfectly consistent with the
field theoretic analysis. Indeed turning on the Romans mass breaks
the trombone symmetry also at the level of the field equations, and
thus it is not consistent to perform the gauging of the trombone
symmetry when the Romans mass parameter is non-vanishing.

\section{Conclusions}
In this paper we have shown that the local $E_{11}$ algebra
corresponding to the IIA theory admits a deformation which is
associated to the gauged IIA theory of \cite{hlw,fibrebundles}. This
deformation is shown to arise from considering the Maurer-Cartan
form that results from taking the eleven-dimensional group element
as in eq. (\ref{tromboneIIAgroupelement}), and solving for the Og
fields using the inverse Higgs mechanism. The deformed algebra can
also be obtained directly in ten dimensions starting from the
commutator of eq. (\ref{deformedinten}) and imposing the closure of
the Jacobi identities, which also imply that this deformation can
not be turned on together with the Romans deformation. Given that
the commutator of eq. (\ref{deformedinten}) involves the trace of
the $GL(10,\mathbb{R})$ generators, this deformation has a
non-trivial effect in the gravity sector, as shown in eqs.
(\ref{deformedspinconnection}) and (\ref{deformedriemanntensor}).

The deformed algebra can naturally be extended to include higher
rank form generators, and we expect the field equations to arise as
duality relations between the corresponding field strengths. It is
important to observe, though, that the only 9-form generator that is
present in the IIA decomposition of $E_{11}$ is associated to the
Romans mass. This can be seen explicitly by observing that the field
strength of the IIA 9-form that one obtains from the deformed
$E_{11}$ algebra associated to the Romans theory
\cite{fabiopeterogievetsky} coincides up to field redefinitions with
the 9-form that one obtains imposing the closure of the
supersymmetry algebra \cite{bkorvp,fabioericIIA}, which also imposes
the duality of its field strength with the Romans mass. Therefore,
there is no dual form in the spectrum associated to the trombone
deformation. In \cite{henning} it was observed that in any dimension
$D$ the $E_{11}$ spectrum contains generators with $D-1$ spacetime
indices in the $(D-2,1)$ mixed symmetry irreducible representation
of $GL(D, \mathbb{R})$ with $D-2$ antisymmetric indices, that could
be associated to the trombone deformations. In this IIA case, this
would be a generator in the $(8,1)$ representation of
$GL(10,\mathbb{R})$, which is indeed present. Actually, the
occurrence of these generators is completely general, as already
shown in \cite{E11dualfields}. Indeed these are the first of an
infinite chain of so called ``dual'' vector generators in the $GL(D,
\mathbb{R})$ representations $(D-2 , D-2 ,..., D-2,1)$, and their
presence is crucial for the universal structure of $E_{11}$
reproducing the gauge algebra of all the form fields in all
dimensions. In \cite{erichalfmax,hierarchythreedimensions} it was
observed that in the case of the internal gaugings one can consider
a lagrangian formulation in which the $D-1$ forms are Lagrange
multipliers for the embedding tensor (so that their field equation
implies the constancy of the embedding tensor). The fact that these
forms are present in the gauge algebra is thus intrinsically related
to the fact that one expects such a lagrangian formulation to be
possible. In \cite{bkorvp} this lagrangian formulation was
originally derived for the IIA case, thus describing simultaneously
the massless and the Romans case. The IIA theory considered in this
paper does not admit a lagrangian formulation, and thus we consider
the fact that there is no form generator associated to this
deformation as completely consistent, and we do not expect any
$E_{11}$ generator associated to a non-propagating field to play a
role in triggering this deformation.

As mentioned in the introduction, in \cite{henning} it was shown
that all possible gauged maximal supergravities of the trombone type
in any dimension $D$ can be classified in terms of a new embedding
tensor in the representation of $E_{11-D}$ which is conjugate to the
one to which the vectors belong. The consistency of the gauge
algebra imposes quadratic constraints, which the authors of
\cite{henning} also analyse in the case in which this trombone
gauging is considered together with the embedding tensor associated
to the internal gauging. In \cite{trombone2} these results are
reproduced imposing the closure of the Jacobi identities of the
deformed local $E_{11}$ algebra with deformations also involving the
trace of the $GL(D, \mathbb{R})$ generators, and the gauge
transformations and the field strengths of the fields are computed
in all cases.

\vskip 2cm

\section*{Acknowledgments}
I would like to thank the organisers of the FPUK v3.0 conference in
Cambridge for creating a stimulating environment while this project
was at its early stages. This work is supported by the PPARC rolling
grant PP/C5071745/1, the EU Marie Curie research training network
grant MRTN-CT-2004-512194 and the STFC rolling grant ST/G000/395/1.

\end{document}